\documentclass[english]{article}
\usepackage[T1]{fontenc}
\usepackage[latin9]{inputenc}
\usepackage[letterpaper]{geometry}
\usepackage{amstext}
\usepackage{esint}
\usepackage{babel}

\begin{document}

\title{\textbf{A (possible) mathematical model to describe biological {}``context-dependence''
: case study with protein structure}}

\author{\textbf{Anirban Banerji}\\
\textbf{Bioinformatics Centre, University of Pune}\\
\textbf{Pune-411007, Maharashtra, India.}\\
\textbf{Email address : anirbanab@gmail.com}}

\date{}
\maketitle
\begin{abstract}
Context-dependent nature of biological phenomena are well documented
in every branch of biology. While there have been few previous attempts
to (implicitly) model various facets of biological context-dependence,
a formal and general mathematical construct to model the wide spectrum
of context-dependence, eludes the students of biology. An objective
and rigorous model, from both 'bottom-up' as well as 'top-down' perspective,
is proposed here to serve as the template to describe the various
kinds of context-dependence that we encounter in different branches
of biology. Interactions between biological contexts was found to
be transitive but non-commutative. It is found that a hierarchical
nature of dependence amongst the biological contexts models the emergent
biological properties efficiently. Reasons for these findings are
provided with a general model to describe biological reality. Scheme
to algorithmically implement the hierarchic structure of organization
of biological contexts was achieved with a construct named 'Context
tree'. A 'Context tree' based analysis of context interactions among
biophysical factors influencing protein structure was performed.\\

\end{abstract}
\\
\textbf{Introduction :}\\
'Context-dependence' is omnipresent in Biology. From the realm
of substitution of nucleotides (Siepel et al. 2004, Zhang et al. 2007)
to the paradigm of protein structure-function (Main et al. 1998, Nobeli
et al. 2009), from the sphere of cellular dynamics (Hagan and Sharrocks
2002) to that in virulence studies in host-parasite systems (Brown
et al., 2003) and evolutionary dynamics (Jablonski et al. 2006), one
encounters events and processes that are {}``context-dependent''.
While various attempts have been made from differing perspectives
to somehow quantify context-sensitiveness of particular biological
events (Andrianantoandro et al. 2006, Torney et al. 2009, Banerji
and Ghosh 2011), a general mathematical framework that attempts to
capture and describe the ubiquitous 'context-dependence', eludes the
students of Biology. In the present work, such a mathematical structure
is proposed that attempts to model biological 'context-dependence'
from bottom-up as well as top-down perspectives. Although the need
to engineer an exact scheme to describe biological context-dependence
was felt by many in recent past (Loewe 2009, Haseltine and Arnold
2007, Marguet et al. 2007, Platzer and Meinzer 2002), the present
approach takes these concerns to a tangible outcome by proposing a
general and robust theoretical construct to model biological organisation
from both top-down and bottom-up perspectives with algorithmically
implementable constructs. Unlike some previous attempts constructs
proposed here do not tangentially touch upon context-dependence modelling
(Standish 2001, Edmonds 1999, Yartseva et al. 2007), but concentrate
solely on it; nor do they restrict themselves into (successful yet)
particular scopes (Doboli et al. 2000, Hoare et al. 2004). On the
other hand, it does not attempt to construct a computational structure
that helps in retrieval of biological data from some repository in
a context-dependent manner (Yu et al. 2009, Boeckmann et al. 2005),
nor does it propose some (effective) visualization tool to observe
context-dependent interactions between biological properties (Gopalacharyulu
et al. 2006). A reliable and general mathematical model to describe
biological context-dependence is of utmost necessity for contemporary
Biology. This paper suggests a possible construct to achieve the same.
(A recent work that attempted constructing a mathematical model to
unambiguously describe the concept 'evolvability' (Valiant 2009),
underlines the necessity of present genre of works.)\\
\\
We suggest the triad of the form <Structure of biological goals$(F)$,
Biological contexts $(C)$, Physical structure of the system $(P)$>,
$\left(<F,C,P>\right)$, to describe the structure of any biological
process. Components of this triad are (of course) not independent;
$F$ determines the suitable choice of $C$; while $C$, in its turn,
operates upon some particular subset of $P$ with definite features.
To ensure $F$ is achieved, $C$ engages only certain elements of
$P$. Such triad-based structure is necessary because same physical
structures can be subjected to different contexts to achieve different
biological goals; for example, same proteins under different set of
contexts may be involved in different biological processes, so that
the goals of these processes (each different) are achieved (Gopalacharyulu
et al. 2006). The present work attempts to formulate general and formal
principles of contextual interactions, by modeling the nature of dependencies
between biological contexts that operate upon physical (structural)
parameters to ensure that biological goals are achieved.\\
\\
Since the motivation of any biological process is solely to accomplish
a set of necessary biological goals, and the structure of biological
goals is hierarchic (Troyanskaya et al. 2003, Camon et al. 2004);
we propose a hierarchic organizational structure for biological contexts
too. Thus, structure of $C$ will assume that of a tree (namely, the
'Context-Tree'($CT$)), where the root-vertex will denote the biological
context necessary to achieve the global goal of the system under consideration.
The lowest level of $CT$ will be occupied by basic, elementary contexts
(- for example, the genes. A systematic combination of the molecular
function of their products (proteins) is studied with respect to achieving
a specific biological goal (Camon et al. 2004)), and their set $A=\left\{ a_{1},a_{2},\ldots,a_{n}\right\} $
will constitute the leaves of $CT$. This form of hierarchic structure
for $CT$, helps in analyzing the measure of performance of any component
of the physical structure of the system under specific context to
achieve any particular $F$, in the form $\;\sum_{i}\frac{\partial F}{\partial a_{i}}\left(=\sum_{i}\frac{\partial F}{\partial C_{i}}.\frac{\partial C_{i}}{\partial a_{i}}\right)$.
In this manner, every context, including the global context, may be
described as a function of composition of basic contextual elements
of $CT$. However, to accomplish a definite set of biological goal,
the contexts describing various facets of a biological system need
to interact between themselves. We define a set of rules $U\left(U=\left\{ \alpha_{1},\alpha_{2},\ldots,\alpha_{n}\right\} \right)$
that governs these interactions. Taken in entirety, they form the
framework for composition between the base elements of $CT$, denoted
by $\alpha$. We note here that set $U$ might be infinite, and the
compositions $a_{i}\alpha a_{j}$ with $\alpha\in U$ need not necessarily
be defined for all $a_{i},a_{j}\in A$ (that is, some of these context
interactions might well be mere theoretical possibilities, unrealized
in biological paradigm).\\
\\
Importance of set $U$ is paramount; it is this set of rules that
governs how the various elements of $CT$ will interact to ensure
the necessary dependencies between contexts, which in turn will (ultimately)
ensure that the system achieves the desired goal $F$. Let us assign
to each element $\alpha$ of the set $U$, an $n\times n$ incidence
matrix (Skiena 1990). $A^{\alpha}$ with entries $a_{ij}^{\alpha}$
is unity if the composition $a_{i}\alpha a_{j}$ is defined, or zero
otherwise. We can then introduce a matrix $A^{U}$ whose elements
are given by :\\
\begin{equation}
a_{ij}=\bigvee_{\alpha\in U}a_{ij}^{\alpha}\end{equation}
\\
stating that $a_{ij}=1$ if there exists a valid composition rule
between $a_{i}$ and $a_{j}$, taken in order.\\
\\
Matrix $A^{U}$ conforms to constraints of biological reality
perfectly because it suggests that for certain magnitudes of $i$,
the $i^{th}$ row and $i^{th}$ column of the matrix $A^{U}$ can
be all zeros. These cases describe algorithmically the fact that interactions
between certain basic elements of context-set $A=\left\{ a_{1},a_{2},\ldots,a_{n}\right\} $
are not allowable biologically. We demonstrate such non-allowable
biological contexts with two examples :\\
\\
\textbf{\underbar{Example-1)}} When $\lambda$ phage (a virus
that infects the bacteria $Escherichia\; coli$) encounters a bacterium,
it attaches itself only to certain particular receptors with specific
structural features, on the bacterial membrane. That is, the relevant
biological contexts ensure that binding of $\lambda$ phage to various
other candidate receptor sites with slightly varying structural aspects,
is not allowed. Subsequently, when the virus genome enters the bacterium,
only two pathways (out of theoretically infinite number of pathways)
of alternative nature, namely the 'lytic pathway' or the 'lysogenic
pathway' are allowed biologically (Yartseva et al. 2007); although
a theoretical thermodynamic study of the situation can suggest many
possible pathways with (almost) similar efficiencies. This entire
process, in the context-space description can be modeled with non-zero
entries for the aforementioned two pathways, while the rest of the
entries in $A^{U}$ will be assigned zero to represent the fact that
$U$ specifies the contexts that ultimately ensures certain biological
goals.\\
\\
\textbf{\underbar{Example-2)}} Out of the entire spectrum of possible
mRNAs that can be generated from a single gene, only one or a few
are created at a time; nature of $A^{U}$ ensures that other possibilities
do not come to being, although these theoretically possible elements
of context-set $A$ would have operated upon the same structural parameters
that constitute$P$. This act of ensuring the interplay between set
of allowable contexts to achieve any particular goal, form the so-called
'regulatory mechanism' that includes an extremely sensitive balance
between concentration magnitudes of pertinent entities, the destination,
sequence variety, structural diversity and finally the functional
options of the resulting protein related to the type of tissue or
the stage of development, etc.. (Boeckmann et al. 2005).\\
\\
Henceforth, we remove corresponding rules to describe the biological
impossibilities in $U$, making it sure thereby that these compositions
do not participate in the construction of the tree $CT$. Removing
these aforementioned entries, from the set $A$ we obtain a non-redundant
set of basic rules of context composition. From here onwards, we will
denote this non-redundant set as set $A$.\\
\\
\textbf{Model :}\\
\textbf{Section-1) : Modeling biological context-dependence from
bottom-up perspective}\\
We can now attempt to describe the biological process of transportation,
as an example. We assume three elements $\left\{ a_{1},a_{2},a_{3}\right\} $
from the set of basic contexts $A$. We denote $a_{1}$ as the context
for loading, $a_{2}$ for delivery and $a_{3}$ for unloading. To
provide an example, we consider the case when axon of a neuron transports
the transmission of action potentials from the cell body to the synapse.
The complicated mechanism of action potential propagation and their
active transportation from their site of synthesis in the cell body
through the axoplasm to intracellular target sites in the axon and
synapse, - provides an ideal case to demonstrate the various context-specific
activities. Here the loading and delivery contexts ($a_{1}$ and $a_{2}$
respectively) involves the series synchronized couplings between allowed
set of specific membranous organelles, synaptic vesicle precursors,
signaling molecules, growth factors, protein complexes, cytoskeletal
components, sodium and potassium channels and many other biological
components. On the other hand, the unloading context ($a_{3}$) describes
the neurotrophic signals that are transported back from the synapse
to the cell body, keeping an account of efficiency and reliability
of loading and delivery operations (Duncan and Goldstein 2006). Here
we introduce the set of compositional rules \textbf{$U$}. $U$ may
then operate upon these three basic elements to ensure a sequential
execution of elementary processes to ensure that final goal ($F$)
is achieved. In this case the matrix $A^{U}$ can be written as :\\
\begin{equation}
\begin{array}{cccc}
 & a_{1} & a_{2} & a_{3}\\
a_{1} & 0 & 0 & 0\\
a_{2} & 1 & 0 & 0\\
a_{3} & 0 & 1 & 0\end{array}\end{equation}
\\
Hence the elementary biological context, $\alpha$, is defined
for only two cases from the entire spectrum of possible contextual
interactions; they are $a_{2}\alpha a_{1}$ (implying the existence
of a pipeline where context for delivery comes into action after the
context for loading is ensured) and $a_{3}\alpha a_{2}$ (implying
the existence of context for unloading after the operation of context
for delivery is performed). It is interesting to note that a pipeline
with $a_{3}\alpha a_{1}$ is not biologically relevant, since it implies
the existence of context for unloading after the context for loading
without any delivery action. Similarly contextual relations like $a_{1}\alpha a_{1}$
(loading certain contexts repeatedly without a purpose), $a_{1}\alpha a_{2}$
(existence of the context to ensure delivery when the initiation of
the process is not ensured), $a_{2}\alpha a_{3}$ (context to ensure
unloading when the delivery process is incomplete) and other spurious
context-relations like $a_{2}\alpha a_{2}$, $a_{3}\alpha a_{3}$
are assigned a magnitude zero because they are biological impertinent.
On a global scale, the composition $a_{3}\alpha a_{2}\alpha a_{1}$
captures the purposeful nature of biological goals.\\
\\
The findings of the last paragraph implies that interactions between
biological contexts, are transitive but are not commutative (since
$a_{3}\alpha a_{2}$ and $a_{2}\alpha a_{1}$ are defined, $a_{3}\alpha a_{2}\alpha a_{1}$
can be defined; but merely because $a_{3}\alpha a_{2}$ exists doesn't
imply that $a_{2}\alpha a_{3}$ exists too). The non-commutative nature
of biological contexts can be understood better from the general treatment
of the problem elaborated later.\\
\\
\\
The (bottom-up) paradigm of description of interplay of biological
contexts can be generalized by describing the entire biological universe
with the Thread-Mesh (TM) model (Banerji 2009). The TM model segments
the biological space-time into a series of different biological organizations,
viz. the nucleotides; amino acids; macromolecules (proteins, sugar
polymers, glycoproteins); biochemical pathways; network of pathways;
biological cell; tissue; organs; organisms; society and ecosystem;
where these organizational schemes are called threshold levels. Emergence
of a single biological property (compositional and/or structural and/or
functional) creates a new biological threshold level in the TM model.
Thus, if any arbitrarily chosen i$^{\text{th}}$ biological threshold
level is denoted as \textbf{TH}$_{i}$, the succeeding one, viz. \textbf{TH$_{\text{i+1}}$}
will be containing at least one biological property that \textbf{TH$_{\text{i}}$}
didn't possess. Schemes with similar philosophy to identify biological
threshold levels were proposed previously (Testa and Kier 2000, Dhar
2007), but representation of emergence of any biological property
and subsequent classification of biological organization with respect
to this emergent behavior, was not done in either of these models.
The basic principles for subsequent discourse are general and can
be applied to any threshold level. Every possible property that a
threshold level is endowed with, is represented by a 'thread' in the
TM model. Thus an environmental property capable of influencing biological
action will be called as an 'environmental thread' in the present
parlance. Threads can be compositional, structural or functional.
For example, for the biological threshold level corresponding to the
enzymes (threshold level representing the macromolecules), one of
the compositional threads is the amino acid sequence; whereas the
radius of gyration, the resultant backbone dipole moment and each
of the bond lengths, bond angles, torsion angles are some examples
of structural threads and the values for K$_{\text{m}}$, V$_{\text{max}}$,
K$_{\text{cat}}$ are some examples of it's functional threads. It
is advantageous to work with the TM model because it can attempt describing
context-dependence and emergence from the framework of an invariant
template.\\
\\
\\
\textbf{Section-2) : Modeling biological context-dependence from
top-down perspective}\\
While framework of $eq^{n}s$ $[1-5]$ along with examples 1 and
2 describe the nature of multilevel organization of CT, such description
is 'bottom-up' in nature. Hence, while it is helpful to describe the
context-mapping between any two particular adjacent biological threshold
levels $'l'$ and $'l+1'$, (say between threshold levels representing
nucleotides and amino acids, amino acids and proteins, or between
proteins and biochemical pathways, etc ..) the general mode of dependency
within CT with a birds-eye ('top-down') view of the organization of
it, can hardly be guessed from such bottom-up approach.\\
\\
We start the construction of the top-down scheme of description
of dependencies between biological contexts, by enlisting the assumptions
involved therein. Hence :\\
\textbf{\underbar{Assumption-1 ) :}} In absence of random external
disturbances and without a failure of any component belonging to physical
structure of the system $(P)$ all the rules of multilevel interactions
between the contexts representing any biological threshold level,
can be constructed and described in deterministic manner. (Success
of recent attempts with deterministic modeling of various biological
phenomena from diverse backgrounds (Janda and Gegina 2008, Kim and
Maly 2009, Ferreira and Azevedo 2007) suggest that such assumption
is not ill-founded, and that too in absence of possible perturbations.)\\
\textbf{\underbar{Assumption-2 ) :}} The necessary and sufficient
condition in order these deterministic rules of inter-level context
interactions hold true, is in their accounting for the accomplishment
of certain biological goals $(F)$. (Previous studies (Yartseva et
al. 2007, Troyanskaya et al. 2003, Camon et al. 2004) vindicate such
assumption.)\\
\textbf{\underbar{Assumption-3 ) :}} Although biological systems
will be exposed to randomly varying magnitudes of external parameters,
the essence of the deterministic criteria of context interactions
in order to accomplish any set of required biological function, will
not be perturbed by significant margin. This assumption implies that
deterministic manner of context interactions will not be undergoing
significant change when the magnitudes of components of underlying
physical structures $\left\{ p_{i}\right\} $ $\left(p_{i}\in P\right)$,
comprised of relevant biological parameters, are altered within some
allowable range. We describe this allowable range of assumed magnitude
of some arbitrarily chosen parameter $\pi$ by an interval $\left[\pi_{0},\pi_{1}\right].$
\\
\\
Relevance of the last assumption is easily understood when one
analyzes the nature of some previous results in depth. Since every
biological property operates within a specified bound of magnitude,
something that has been referred to as 'fluctuation' in an earlier
study (Testa and Kier 2000) the functions that represent them will
also be bounded in their ranges. Examples for such fluctuation are
many; in the biological threshold level representing cells, for the
mitogen-activated protein kinase cascade studies, the total concentrations
of MKKK, MKK and MAPK have been found to be in the range 10\textendash{}1000
nm and the estimates for the K$_{\text{cat}}$ values of the protein
kinases and phosphatases have been found to range from 0.01 to 1 s$^{\text{-1}}$(Kholodenko
2000). Similarly, for\textbf{ }the proteins, the mass fractal dimension
and hydrophobicity fractal dimension representing compactness of mass
and hydrophobicity distribution, have been found to be in the range
between 2.18 to 2.37 and 2.22 to 2.43 respectively (Banerji and Ghosh
2009). \\
\\
Based on these assumptions, we propose the functional that defines
the probability of attaining the biological goal $(F)$ under consideration,
to assume the form :\\
\begin{equation}
F_{0}=\int_{x_{i}\in S}\phi\left(x_{1},x_{2},\ldots,x_{n}\right)dx_{i}\;\;\;\;\;\;\;\;\left(1\leq i\leq n\right)\end{equation}
\\
where $\phi$ is the probability density of attaining the objective
(biological goal) and $X$ is the feasibility domain of the contexts
$x_{i}$. \\
\\
\\
Since to achieve every biological goal, many (say, $m$) successive
stages of context interactions are required, we can express $eq^{n}-6$
at a higher resolution as :\\
\begin{equation}
\phi\left(x_{1},x_{2},\ldots,x_{n}\right)=\prod_{j=1}^{m}\phi_{j}|\phi_{j-1}\left(x_{1},x_{2},\ldots,x_{n}\right)\end{equation}
\\
where $\phi_{j}|\phi_{j-1}$ represent the conditional probability
associated with context $\left\{ x_{i}\right\} $ interactions, while
attempting to achieve a particular biological goal.\\
\\
However, we note that individual physical parameters $p_{i}$
$\left(p_{i}\in P\right)$, upon which the contexts are working, may
not always be strongly correlated and although related with each other,
can be considered independent when viewed individually with respect
to their functional contribution to the system. For example, the time-dependent
and context-dependent fluctuations in individual bond lengths, bond
angles and torsion angles in the protein interior, although might
be related in some intricate way to the resultant dipole moment for
the protein; can be considered, for all practical purposes, in terms
of their individual (and not linked) contributions in ensuring proteins
stability and functionality. Hence we attempt to partition the relevant
contexts into a sum of disjoint domains; such that :\\
$\left(x_{i}\in X_{i}\right)$ and $\sum_{i}X_{i}=S$.\\
\\
Considering this partition we can re-write $eq^{n}-6$ as :\\
\begin{eqnarray}
F_{0}=\int_{x_{1}\in X_{1}}\int_{x_{2}\in X_{2}}..\int_{x_{n}\in X_{n}}\phi_{1}\left(x_{1},x_{2},..,x_{n}\right)\times\phi_{2|1}\left(x_{1},x_{2},..,x_{n}\right)..\phi_{m|m-1}\left(x_{1},x_{2},..,x_{n}\right)dx_{1}dx_{2}..dx_{m}\end{eqnarray}
\\
In other words, purely in terms of achievement of biological goals
:\\
\begin{equation}
F_{0}=F_{1}F_{2|1}\ldots F_{m|m-1}\end{equation}
\\
where\\
\begin{equation}
F_{j|j-1}=\int_{x_{1}\in X_{1}}\int_{x_{2}\in X_{2}}\ldots\int_{x_{n}\in X_{n}}\phi_{j|j-1}\left(x_{1},x_{2},\ldots,x_{n}\right)dx_{1}dx_{2}\ldots dx_{n}\end{equation}
\\
are the conditional probabilities of context-interactions of the
system realizing the successive stages of the task. It is necessary
to mention here that to achieve any biological function, the domain
of integration for every $x_{i}$ in $eq^{n}-8$ must be within their
respective permissible range, say $\left[x_{\pi_{0}},\;\; x_{\pi_{1}}\right]$.\\
\\
While it is difficult to assume that every context-interaction
necessary to realize certain biological goal will always be operating
in deterministic manner with perfect efficiency, experience teaches
us that biological goals are seldom compromised with. Hence we assume
that the reliability of any arbitrarily chosen context interaction
at any arbitrarily chosen $j^{th}$ state in the realization of certain
biological function, is statistically independent of the probability
of the realization of that particular biological function. In that
case, the integrand in $eq^{n}-10$, can be expressed as a product
$\phi_{j|j-1}\left(x_{1},x_{2},\ldots,x_{n}\right)r_{j}\left(x_{1},x_{2},\ldots,x_{n}\right)$,
where $r_{j}$,$\left(r_{j}\in R\right)$ describes the probability
of reliability of any arbitrarily chosen context-interaction at $j^{th}$
state in the realization of certain biological function.\\
\\
Hence $eq^{n}-7$, can be expressed more realistically as :\\
\begin{equation}
\phi\left(x_{1},x_{2},\ldots,x_{n}\right)=\prod_{j=1}^{m}\phi_{j|j-1}\left(x_{1},x_{2},\ldots,x_{n}\right)r_{j}\left(x_{1},x_{2},\ldots,x_{n}\right)\end{equation}
\\
\\
Thus, when the reliability of context-interactions are taken into
account, assuming that $eq^{n}-8$ and $eq^{n}-10$ are valid, $eq^{n}-6$
and $eq^{n}-7$ can be re-written as :\\
\begin{equation}
F_{0}=\prod_{j=1}^{m}F_{j}|F_{j-1}R_{j}\end{equation}
\\
where\\
\begin{equation}
R_{j}=\int_{x_{1}\in X_{1}}\int_{x_{2}\in X_{2}}\ldots\int_{x_{n}\in X_{n}}\;\; r_{j}\left(x_{1},x_{2},\ldots,x_{n}\right)dx_{1}dx_{2}\ldots dx_{n}\end{equation}
\\
In other words, $eq^{n}-12$, can be re-written as :\\
\begin{equation}
F_{0}=F_{1}F_{2|1}\ldots F_{n|n-1}R_{1}R_{2}\ldots R_{m}\end{equation}
\\
\textbf{Result :}\\
\textbf{(Bottom-Up) Modeling hierarchical organization with 'context
tree':}\\
\textbf{Case-study with protein structure :}\\
To describe the 'Context-Tree'($CT$) in such hierarchic paradigm
under a generalized scheme we introduce a construct $C$, which is
a family of embedded partitions of contexts $C=<C^{1},C^{2},\ldots,C^{r}>$
that operate upon any relevant subset of structural threads $(J)$
representing the physical structure $(P)$ of any arbitrarily chosen
threshold level $S$. $J\subset P$ and $J=\left\{ 1,2,\ldots,m\right\} $.
For example, it has been found (Main et al. 1998) that at the threshold
level of proteins; $\left(S\,:\:\mathbf{TH_{Proteins}}\right)$ in
an urea-induced media (the 'environmental thread' influencing $J$),
the extent of stability of mutant proteins are highly dependent on
the contexts $(C)$ which operate upon the various structural parameters
$(J)$, that form a subset of $(P)$ describing $(S)$.\\
\\
We can describe the situation as :\\
\begin{equation}
C^{S}=<C_{1}^{S},C_{2}^{S},\ldots,C_{l}^{S}>\quad\cup_{j=1}^{l}C_{j}^{S}=J,\quad C_{i}^{S}\cap C_{j}^{S}=\emptyset\:\:\left(i\neq j\right),\;\: S=\overline{1,\: r}\end{equation}
\\
The embedding refers to any element of the partition of the $S^{th}$
biological threshold level; i.e., the set $C_{j}^{s}$ represents
the union of several sets $C_{i_{1}}^{S-1},C_{i_{2}}^{S-1},\ldots,C_{i_{z}}^{S-1}$
of the $\left(S-1\right)^{th}$ biological threshold level. Such description
of ($CT$) conforms to a previous study on similar topic (Andrianantoandro
et al. 2006). Findings from a recent study (Haseltine and Arnold 2007)
vindicates $C_{i}^{S}\cap C_{j}^{S}=\emptyset$. To elaborate the
hierarchic structure, we can write $C\longleftrightarrow<C_{1}^{S-1},C_{2}^{S-1},\ldots,C_{l}^{S-1}>$,
if $C^{S}=\cup_{i=1}^{l}C_{i}^{S-1}$. Since the entire set of interactions
between various contexts is ultimately geared to satisfy biological
goals and since the nature of organization of biological goals is
hierarchic, we attempt to describe it by defining $\: l\left(C^{S}\right)=l$
and $C^{root}=<\left\{ 1,2,\ldots,m\right\} >$ ; i.e., the partition
at the highest (root) level consists of one set, namely $J$.\\
\\
\\
We can associate each element $C_{j}^{S}\;\:\left(s=\overline{2,\: r}\right)$
of the partition to the context-interaction function, namely $f_{j}^{S}\left(\alpha_{1},\alpha_{2},\ldots,\alpha_{l\left(C_{j}^{S}\right)}\right)$,
where $\alpha\in\left\{ 0,-1,+1\right\} $, conforming to the previously
defined $U\left(U=\left\{ \alpha_{1},\alpha_{2},\ldots,\alpha_{n}\right\} \right)$.
We associate each element $C_{j}^{1}$ of the first level to a binary
relation $R_{j}$(the previously defined elementary contexts are related
by this, say $aRb$, where $A=\left\{ a,b,\ldots,z\right\} $) on
the biological sub-space $EC_{j}^{1}$. \\
\\
These concepts can formally be described as :\\
Let $C_{j}^{2}\longleftrightarrow<C_{1}^{1},\ldots,C_{l}^{1}>$
and define a relation $R_{j}^{2}$ on $E_{C_{j}^{2}}$ using the formula
(for $l>1$) :\\
\begin{equation}
a_{C_{j}^{2}}R_{bC_{j}^{2}}^{2}\Longleftrightarrow f_{j}^{2}\left(\alpha_{1},\alpha_{2},\ldots,\alpha_{l}\right)=1\end{equation}
\\
\begin{eqnarray*}
\alpha_{i} & = & +1\quad\: if\;\, a_{C_{i}^{1}}R_{i}b_{C_{i}^{1}}\\
\alpha_{i} & = & -1\quad\: if\;\, a_{C_{i}^{1}}\overline{R}b_{C_{i}^{1}}\\
\alpha_{i} & = & 0\quad\: if\;\, a_{C_{i}^{1}}=b_{C_{i}^{1}}\end{eqnarray*}
\\
In case of $l=1$, $R_{j}^{2}=R_{j}$. If all the relations $R^{S-1}$
of the $\left(S-1\right)^{th}$ biological threshold level are defined,
then the relations $R_{j}^{S}$ of the $S^{th}$ threshold level with
$\left(l>1\right)$ can be defined by the following construct :\\
\\
If $C_{j}^{S}\longleftrightarrow<C_{1}^{S-1},C_{2}^{S-1},\ldots,C_{l}^{S-1},>$
, then\\
\begin{equation}
a_{C_{j}^{S}}R_{bC_{j}^{S}}^{S}\Longleftrightarrow f_{j}^{S}\left(\alpha_{1},\alpha_{2},\ldots,\alpha_{l}\right)=1\end{equation}
\\
\begin{eqnarray*}
\alpha_{i} & = & +1\quad\: if\;\, a_{C_{i}^{S-1}}R_{i}^{S-1}b_{C_{i}^{S-1}}\\
\alpha_{i} & = & -1\quad\: if\;\, a_{C_{i}^{S-1}}\overline{R^{S-1}}b_{C_{i}^{S-1}}\\
\alpha_{i} & = & 0\quad\: if\;\, a_{C_{i}^{S-1}}=b_{C_{i}^{S-1}}\end{eqnarray*}
\\
If $C^{S}=C^{S-1}$, then $R^{S}=R^{S-1}$; in other words the
construction requires the relation $R$ to coincide with $R^{r}$,
which is a single relation at the $r^{th}$ upper level.\\
\\
To describe the entire bottom-up paradigm of description of interaction
scheme between biological contexts, we consider an example where we
describe the contextual constraints on the active site of an enzyme
in simplistic terms. For this case, without any loss of generality,
we consider the threshold level representing proteins to be the root
level in this case. The goal of the system $\left(F\right)$ for the
present purpose is to make the enzyme functional. We assume the elementary
contexts that can influence functionality of the enzyme active site
to be represented with 3 basic partitions; namely, first, the contextual
differences originating out of internal coordinates; second, contextual
differences arising out of interaction profile of the active site
atoms with water; and third, contextual differences arising out of
the capability of the active site to undergo a shape change. Hence
we describe the family of partitions $C$, as $C^{1}=<\left\{ 1,2,3\right\} ,\;\left\{ 4,5\right\} ,\;\left\{ 6\right\} >$
and $C^{2}=<J>$; where element 1 denotes (possible) contextual difference
arising out of the fluctuation of bond lengths, element 2 denotes
(possible) contextual difference arising out of the fluctuation of
bond angles, element 3 denotes (possible) contextual difference arising
out of the fluctuation of torsion angles. Similarly, element 4 stands
for (possible) contextual difference arising out of the hydrophobicity
of active site patch, element 5 denotes the (possible) contextual
difference arising out of the local electrostatic profile of the active
site patch. Element 6 denotes the extent of (possible) contextual
difference arising out of the change in the local shape of the active
site patch. Denoting the set $\left\{ 1,2,3\right\} $ as $D_{1}^{2}$,
$\left\{ 4,5\right\} $ as $D_{2}^{2}$ and $\left\{ 6\right\} $
as $D_{3}^{2}$, the hierarchic nature of these contextual dependencies
can easily be described as :\\
\\
$D_{2}RD_{1}\Longleftrightarrow\left[D_{2}\geq D_{1}\right]$\\
similarly, \\
$D_{3}RD_{2}\Longleftrightarrow\left[D_{3}\geq D_{2}\right]$
and $D_{3}RD_{1}\Longleftrightarrow\left[D_{3}\geq D_{1}\right]$
\\
\\
Implying that a (possible) contextual difference arising out of
the hydrophobicity of active site patch, or a (possible) contextual
difference due to the local electrostatic profile of the active site
patch will surely account for some change in the distribution profile
of bond length, bond angle and torsion angle distribution. But inverse
of this case, viz., a (secondary) change in the local electrostatics
profile and local hydrophobic profile due to a (primary) change in
bond-length, bond-angle or torsion angle might or might not be observed
in reality. Similarly, in case of a possible change in local shape
the local electrostatic profile, local hydrophobicity profile, local
distribution of bond length, bond angle, torsion angle will surely
be taking place; but the other way round might or might not be observed.
This vindicates and generalizes our previous finding that interactions
between biological contexts, are transitive but are not commutative
(hence, if $D_{2}RD_{1}$ and $D_{3}RD_{2}$ are defined, $D_{3}RD_{2}RD_{1}$
can be defined; but merely because $D_{3}RD_{2}$ exists doesn't imply
that $D_{2}RD_{3}$ exists too).\\
\\
\textbf{\underbar{Conclusion :}}\\
While $eq^{n}-1$ to $eq^{n}-5$ constructed the bottom-up scheme
of describing the interactions and dependencies between biological
contexts, the framework of equations $eq^{n}-6$ to $eq^{n}-14$ describe
the top-down view of the same. Together these set of equations present
a comprehensive way to quantitatively model the omnipresent {}``context-dependence''
in biology. Evidences for the reliability of such mathematical treatise
can easily be obtained from the various experimentally proved results
that are provided to emphasize the reasonable nature of these formulations.
Since contemporary biology, as never before, is attempting to be objective
in its philosophy, the necessity of a mathematical model to describe
the {}``context-dependent'' nature of it can hardly be ignored.
The model proposed here, therefore, assumes immense importance.\\
\\
\textbf{\underbar{Acknowledgment}}\underbar{ }\textbf{\underbar{:}}
This work was supported by COE-DBT (Department of Biotechnology, Government
of India) scholarship.\\
\\
\textbf{\underbar{References :}}\\
Andrianantoandro E, Basu S, Karig D, Weiss R (2006) Synthetic
biology: new engineering rules for an emerging discipline. Mol. Syst.
Biol. 2:0028.\\
Banerji A, Ghosh I (2009) Revisiting the Myths of Protein Interior:
Studying Proteins with Mass-Fractal Hydrophobicity-Fractal and Polarizability-Fractal
Dimensions. PLoS One 4(10), e7361.\\
doi:10.1371/journal.pone.0007361.\\
Banerji A (2009) Existence of biological uncertainty principle
implies that we can never find 'THE' measure for biological complexity.
arXiv:0902.0490v2.{[}q-bio.OT{]}\\
Banerji A, Ghosh I (2011) Mathematical criteria to observe mesoscopic
emergence of protein biochemical properties. J. Math. Chem. 49(3):643-665.\\
Boeckmann B, Blatter M, Famiglietti L, Hinz U, Lane L, Roechert
B, Bairoch A (2005) Protein variety and functional diversity: Swiss-Prot
annotation in its biological context. C. R. Biologies 328,882-899.\\
Brown M, Schmid-Hempel R, Schmid-Hempel P (2003) Strong context-dependent
virulence in a host-parasite system: reconciling genetic evidence
with theory. J. Anim. Ecol., 72(6):994-1002.\\
Camon E, Magrane M, Barrell D, Lee V, Dimmer E, Maslen J, Binns
D, Harte N, Lopez R, Apweiler R (2004) The gene ontology annotation
(goa) database: sharing knowledge in uniprot with gene ontology. Nucl.
Acids Res. 32:D262-D266.\\
Dhar P (2007) The next step in biology: A periodic table? J. Biosci.
32:1005\textendash{}1008.\\
Doboli S, Minai A, Best P (2000) Latent attractors: a model for
context-dependence place representations in the hippocampus, Neural
Comput. 12(5): 1009-1043.\\
Duncan J, Goldstein L (2006) The Genetics of Axonal Transport
and Axonal Transport Disorders; PLoS Genet. 2(9): e124.\\
Edmonds B (1999) Syntactic measures of complexity. Ph.D. thesis,University
of Manchester.\\
http://www.cpm.mmu.ac.uk/\textasciitilde{}bruce/thesis\\
Ferreira P, Azevedo P (2007) Evaluating deterministic motif significance
measures in protein databases; Algorithms for Mol. Biol., 2:16.\\
Gopalacharyulu P, Lindfors E, Bounsaythip C, Oresic M (2006) Context
dependent visualization of protein function, In: Rousu J, Kaski S,
Ukkonen E (eds) Proc. Probabilistic Modelling and Machine Learning
in Structural and Systems Biology Workshop, pp 26-31.\\
Hagan I, Sharrocks A (2002) Understanding cancer: from the gene
to the organism. Conference on genes and cancer. EMBO Rep. 3(5):415\textendash{}419.\\
Haseltine E, Arnold F (2007) Synthetic Gene Circuits: Design with
Directed Evolution. Ann. Rev. Biophys. Biomol. Struct. 36:1-19.\\
Hoare D, Couzin I, Godin J, Krause J (2004) Context-dependent
group size choice in fish. Anim. Behav. 67:155-164.\\
Jablonski P, Lee S, Jerzak L (2006) Innate plasticity of a predatory
behavior: nonlearned context dependence of avian flush-displays. Behav.
Ecol. 17(6): 925-932.\\
Janda J, Gegina G (2008) A deterministic model for the processing
and presentation of bacteria-derived antigenic peptides. J. Th. Biol.
250(3): 532-546.\\
Kholodenko B (2000) Negative feedback and ultrasensitivity can
bring about oscillations in the mitogen-activated protein kinase cascades.
Eur. J. Biochem. 267(6):1583\textendash{}1588.\\
Kim M, Maly I (2009) Deterministic mechanical model of T-killer
cell polarization reproduces the wandering of aim between simultaneously
engaged targets. PLoS Comp .Biol. 5(1):1-12.\\
Loewe L (2009) A framework for evolutionary systems biology, BMC
Sys. Biol. 3:27.\\
Main E, Fulton K, Jackson S (1998) Context-Dependent Nature of
Destabilizing Mutations on the Stability of FKBP12. Biochemistry.
37(17):6145\textendash{}6153.\\
Marguet P, Balagadde F, Tan C, You L (2007) Biology by design:
reduction and synthesis of cellular components and behaviour. J. R.
Soc. Interface, 4:607-623.\\
Nobeli I, Favia A, Thornton J (2009) Protein promiscuity and its
implications for biotechnology. Nature biotechnology; 27(2):157-167.\\
Platzer U, Meinzer H (2002) Simulation of genetic networks in
multicellular context.\\
In: Polani D, Kim J, Martinez T (eds) 5th German workshop on artificial
life: abstracting and synthesizing the principles of living systems,
Berlin, Germany: Akad. Verl.-Ges, pp. 43-51.\\
Siepel A, Haussler D (2004) Phylogenetic Estimation of Context-Dependent
Substitution Rates by Maximum Likelihood. Mol. Biol. Evol. 21(3):468-488.\\
Skiena S (1990) Implementing Discrete Mathematics: Combinatorics
and Graph Theory with Mathematica. Reading, MA: Addison-Wesley, pp.
135-136.\\
Standish R (2001) On Complexity and Emergence, arXiv:nlin/0101006v1
{[}nlin.AO{]}.\\
Testa B, Kier L (2000) Emergence and Dissolvence in the Self-organisation
of Complex Systems. Entropy 2:1-25.\\
Torney C, Neufeld Z, Couzin I (2009) Context-dependent interaction
leads to emergent search behavior in social aggregates. Proc. Nat.
Acad. Sc. USA, 106(52):22055-22060.\\
Troyanskaya O, Dolinski K, Owen A, Altman R, Botstein D (2003)
A Bayesian framework for combining heterogeneous data sources for
gene function prediction (in Saccharomyces cerevisiae). Proc. Nat.
Acad. Sc. USA 100(14):8348 \textendash{} 8353.\\
Valiant L (2009) Evolvability. J. Assoc. Comp. Mach. 56,1, 56:1,
3:1-3:21.\\
Yartseva A, Klaudel H, Devillers R, Kepes F (2007) Incremental
and unifying modelling formalism for biological interaction networks.
BMC Bioinformatics 8: 433.\\
Yu C, Zavaljevski N, Desai V, Reifman J (2009) Genome-wide enzyme
annotation with precision control: catalytic families (CatFam) databases.
Proteins. 74:449-460.\\
Zhang W, Bouffard G, Wallace S, Bond J (2007) Estimation of DNA
Sequence Context-dependent Mutation Rates Using Primate Genomic Sequences.
J. Mol. Evol. 65:207-214. \\

\end{document}